\documentclass{article}
\usepackage{hiph-preprint,epsfig}
\volnumber{22} \issuenumber{1} \edyear{2005}                             
\frompage{000} \topage{000}                                              
\recrevdate{31 October 2005}                                              
\def\rightmark{Uncertainties in Air Showers from Small-x pQCD Mini-Jets}
\def\leftmark{L. Portugal}

\newcommand{\be}{\begin{equation}}
\newcommand{\ee}{\end{equation}}
\newcommand{\bea}{\begin{eqnarray}}
\newcommand{\eea}{\end{eqnarray}}
\newcommand{\gsim}{\mbox{\raisebox{-0.6ex}{$\stackrel{>}{\sim}$}}\:}

\title{Uncertainties in Air Showers from Small-x pQCD Mini-Jets}

\authors{ 
{L. Portugal$^{1,2}$ %
\index{One, A.} 
\index{Two, A.} 
}\\[2.812mm]
{\normalsize
\hspace*{-8pt}$^1$ Institut f\"ur Theoretische Physik, 
J.W.\ Goethe Universit\"at \\ 
Max-von-Laue Str.\ 1, 60438 Frankfurt, Germany\\[0.2ex] 
\hspace*{-8pt}$^2$ Instituto de Fisica, Universidade Federal do Rio de Janeiro\\ 
Rio de Janeiro, Brazil
}}

\abstract{
To investigate uncertainties 
in air shower simulations caused by the small-$x$ regime,
a model including leading-twist hard pQCD plus soft
processes was built, which are separated by an energy dependent
transverse momentum cut-off. We provide a fit of the cut-off
to the total $pp$ cross section for different PDFs using the eikonal
formalism and show that for modern PDF sets there is only a small
uncertainty in the mini-jet cross section, and hence in the final-state
multiplicity and the number of produced muons.}



\makeindex
\begin{document}
\maketitle

\section{Introduction}

Extensive air showers (EAS), are induced in the atmosphere due to
collisons of a primary cosmic ray particle with air nuclei.  The
properties of cosmic rays, such as their origin, energy and
composition are (partly) reconstructed from air shower
observables. Thus, understanding the physics of the primary collision
and of subsequent interactions in the atmosphere is
important. The main uncertainties are dominated
by hadronic interactions. For example, muon multiplicities at Auger, Kascade
and CosmoLEP \cite{muon} might hint at a poor understanding of
particle production at ultra-high cosmic ray energies.  Given that
these extend far beyond those of terrestrial accelerators, our
knowledge of hadronic interactions has to be extrapolated to unknown
regimes in any QCD inspired model attempting to describe the physics
of EAS~\cite{engel}.

The QCD mini-jet cross section in $pp$ increases rapidly with energy
and dominates the total cross section already for collider
energies. It is sensitive to the small-$x$ regime of perturbative QCD
(pQCD), and to the break-down of the leading-twist approximation
at small transverse momentum, which requires us to introduce a
cut-off ${p_T}_{cut}$.

Another source of uncertainties is low-momentum ``soft'' particle
production which dominates the very forward rapidity region and hence the
momentum degradation of the cosmic ray in the atmosphere. For
semi-central collisions with air nuclei an IR-safe approach to
particle production is provided by gluon saturation effects,
see~\cite{DDS} for an application to EAS. For peripheral $pA$ and for
$pp$ collisions no intrinsic semi-hard scale exists and weak-coupling
methods are therefore not available; here, we followed instead the
Dual Parton Model~\cite{DPM}, adjusting phenomenological parameters to
(recent) accelerator data.  However, our present focus is on
uncertainties related to the perturbative QCD regime, in particular
with respect to the multiplicity of secondaries (or $d\sigma_{jet}/dy$)
which is related to the number of produced muons.

\section{Mini-jet production}\label{model}

The hard jet cross section in pQCD can be written as~\cite{owens}
\be \label{jet_croxsec}
\frac{d \sigma_{jet}}{dy_3 dy_4 d^2 \mathbf{p_T} d^2 \mathbf{b}}=
\sum_{i,j,l,k}
x_1f_i(x_1,Q^2,b)x_2f_j(x_2,Q^2,b)\frac{1}{\pi}
\frac{d\sigma}{d\hat{t}}(i+j \rightarrow l+k),
\ee 
where the summation runs over all parton species, $x_1$ and $x_2$ 
are the light-cone momentum fractions carried by the scattering
partons, $\hat{t}$ is the 
momentum transfer and $\mathbf{b}$ is the impact parameter.
The kinematical variables $x_1,x_2$ and $\hat{t}$ are related in a simple way
to the rapidities $y_3,y_4$ and $\mathbf{p_T}$~\cite{owens}.
We do not employ a phenomenological $K$-factor since its energy
dependence is not well known; this uncertainty does not affect our
results significantly except for increasing the $p_t$-cutoff shown below
slightly.

The functions $f(x,Q^2,b)$ are the PDFs. The impact parameter
dependence is assumed to factorize, $f(x,Q^2,b) =
T(b)f(x,Q^2)$, with $T(b)$ the proton overlap function chosen according 
to~\cite{overlapfunction}. HERA has measured the PDFs over a broad
range of $Q^2=p_T^2$, but only for $x\gsim10^{-4}$. 
Modern PDF packages extrapolate to lower $x$ according to QCD
evolution equations. We shall show results for both the (leading order) GRV98
and CTEQ6 parameterizations~\cite{pdfs} which feature a BFKL-like growth
of the gluon distribution at small $x$. To illustrate the effect of small-$x$
evolution, we also employ the old Duke-Owens (DO) parameterization, which
exhibits a much flatter $xg(x,Q^2)$ at small $x$.
 
Due to the power-law divergence of~(\ref{jet_croxsec}) as $p_T\to0$ we
need to introduce a low momentum cut-off. The (phenomenological) soft
cross section $\sigma_{soft}=57$~mb is assumed to be constant at high
energies. It's value is chosen to reproduce the total $pp$ cross section
at the lower end of relevant energies, $E_{cm}\sim100$~GeV.

In order to avoid partial wave unitarity violation we use the eikonal 
formalism~\cite{eikonal} to define the total cross section. In this 
formalism the total and inelastic cross sections are related 
to $\sigma_{jet}+\sigma_{soft}$ by
\be \label{tot}
\sigma_{tot} = 2 \int d^2{\bf b} [1-e^{-T(b)
    \left(\sigma_{jet}+\sigma_{soft} \right)}]~,~
\sigma_{el} = \int d^2{\bf b} [1-e^{-T(b)
    \left(\sigma_{jet}+\sigma_{soft} \right)}]^2,
\ee
and $\sigma_{inel}=\sigma_{tot}-\sigma_{el}$.

\section{Results} \label{results}

According to {\bf S}-Matrix and Regge theory the total $pp$ cross
section can be parameterized in a variety of ways at high energies.  A
parameterization fitted to the available experimental data is
presented in~\cite{pdg}. Using expression~(\ref{tot}) we fitted
${p_T}_{cut}$ to that parameterization of $\sigma_{tot}$ (for $pp$)
for different PDFs. As shown in fig.~1a, for modern PDFs
the cut-off increases very rapidly at high energies, which is related
to the small-$x$ growth of the gluon distribution. The fits for CTEQ6
and GRV98 return very similar values for ${p_T}_{cut}$, with small
discrepancies only for $E_{cm}\simeq100$~TeV. On the other hand, the
DO set requires a {\em constant} cut-off within the relevant energy
regime, which is due to the too flat small-$x$ gluon distribution for
this old parameterization. 

Using the above ${p_T}_{cut}$ for a given energy and PDF,
we can now integrate~(\ref{jet_croxsec}) over
$y_4$ and $p_T$ to obtain $d \sigma_{jet} /dy$. In fig.~1b we show 
$d \sigma_{jet}/dy$ for different PDFs at $E_{CM}= 150$~TeV, a typical 
primary high-energy cosmic ray. Again we observe very good agreement
among the modern PDFs, both of which predict a much higher multiplicity at
central rapidity than the old DO set. Also, the latter gives a much
wider rapidity plateau, implying a reduced momentum degradation in the
atmosphere from inelastic interactions.

\begin{figure}[!ht]
\begin{center}
\vspace*{-1.5cm}
\begin{tabular}{cc}
 \epsfig{file=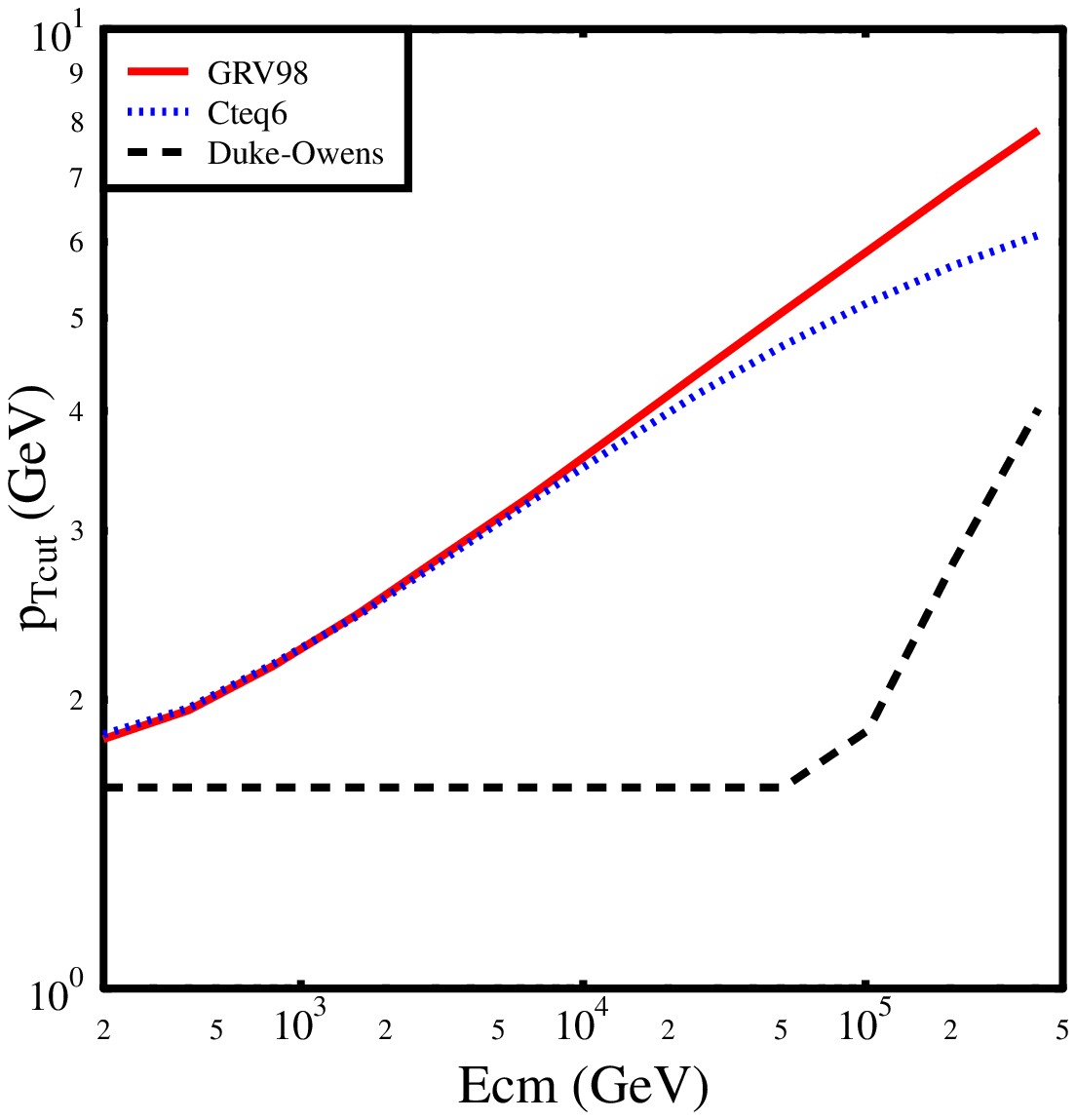,width=6cm}&
 \epsfig{file=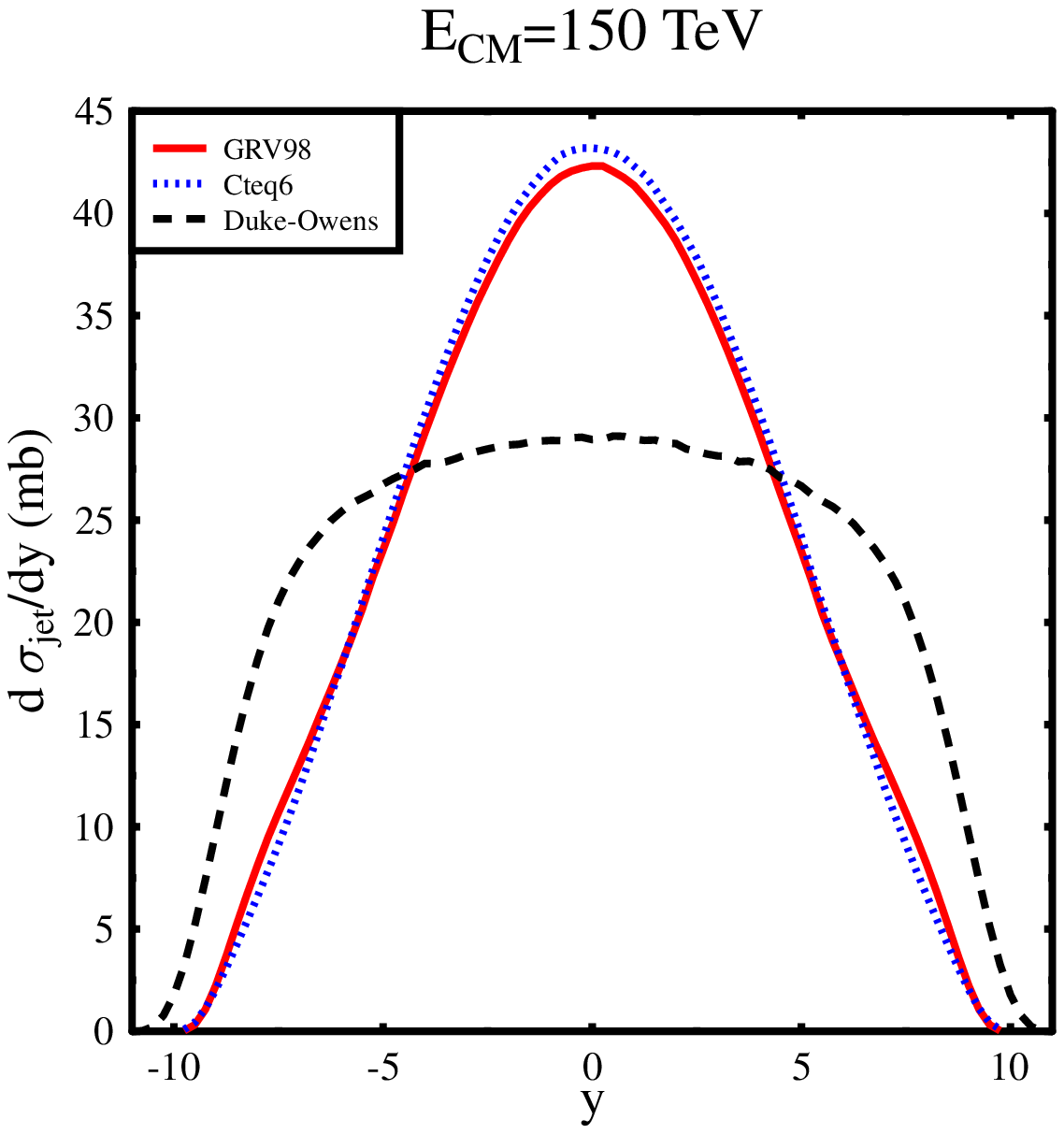,width=6cm}\\
 (a) & (b)  \\
\end{tabular}
\end{center}
\vspace*{-.8cm}
\caption{Fig {\it a)} Transverse momentum cut-off for
  leading-twist LO pQCD processes in $pp$, for
  various sets of PDFs.
 Fig {\it b)} Rapidity distribution of the (mini-) jet cross section
  in inelastic $pp$ collisions for those PDFs.}
\label{fig1}
\end{figure}



\section{Conclusion}

We have studied possible uncertainties in hard process
extrapolated to high energies, which are sensitive to the small-$x$
regime of the PDFs. We fixed ${p_T}_{cut}(E_{cm})$ by fitting the
total $pp$ cross section in the eikonal formalism. 
While modern sets of PDFs with a steep increase of the gluon
distribution at small $x$ agree rather well, old parameterizations
predict a much flatter ${p_T}_{cut}$. 

We have also shown $d \sigma_{jet}/dy$ for different PDFs. There is
little difference between CTEQ6 and GRV98 both at central and forward
rapidity. Since mini-jets dominate particle production if the soft
cross section is energy independent, we do not expect a large
uncertainty in the number of produced muons (approximately
proportional to the multiplicity of secondary hadrons) arising from
the high-energy part of an EAS. For DO, wich exhibts a much slower growth 
of the small $x$ gluon density one obtains a significantly wider rapidity
plateu wich would affect EAS. In the future, we plan to extend our
model to $pA$ collisions and implement it into an EAS simulation
package to study directly the effects of small-$x$ QCD interactions on
air shower observables.

\section*{Acknowledgment(s)}
It is a pleasure to thank the organizers of "Quark matter 2005" for
the opportunity to present this work and H.J.\ Drescher, A.\ Dumitru,
Y.\ Nara and T.\ Kodama for their help with this project. Support by
CAPES and CNPQ is acknowledged.

\vfill\eject
\end{document}